\documentstyle[12pt,emlines]{article}

\def\p{\bot}
\def\pa{\scriptscriptstyle \|}
\def\dd{\partial}
\def\k{\char'32}
\def\d{\delta}
\def\o{\omega}
\def\e{\varepsilon}
\def\f{\varphi}
\def\({\left(}
\def\[{\left[}
\def\){\right)}
\def\]{\right]}
\def\g{\gamma}
\def\a{\alpha}
\def\b{\beta}
\def\t{\tau}
\def\n{\nu}
\def\m{\mu}
\def\s{\sigma}
\def\L{\Lambda}

\newcounter{form}
\newcommand{\dis}[1]{$$\displaylines{#1}$$}
\newcommand{\disn}[2]{$$\displaylines{\refstepcounter{form}
            \label{#1} \hfill #2}$$}
\newcommand{\no}{\hfill \phantom{(\theform)}\cr \hfill}
\newcommand{\nom}{\hfill (\theform) \cr}

\newcounter{punkt}
\makeatletter
\renewcommand{\section}{\@startsection{section}{1}{0pt}%
            {3.5ex plus 1ex minus .2ex}{2.3ex plus .2ex}{\bf}}
\makeatother
\newcommand{\sect}[2]{\protect\refstepcounter{punkt}\protect\label{#1}
            \section*{$\protect\vphantom{a}$\hfill
            \arabic{punkt}.\hskip 2mm #2 \hfill $\protect\vphantom{a}$}}
\newcommand{\st}{\hfill $\protect\vphantom{a}$\protect\\
            $\protect\vphantom{a}$\hfill}

\begin{document}

\title{Comparison of quantum field perturbation theory \\
       for the light front with the theory \\ in lorentz coordinates}

\author{S. A. Paston\thanks{E-mail: Sergey.Paston@pobox.spbu.ru},
V. A. Franke\thanks{E-mail: franke@snoopy.phys.spbu.ru}\\
St.-Peterspurg State University, Russia}

\date{24 January 1997}

\maketitle
\begin{abstract}

The relationship between the perturbation theory in light-front
coordinates and \linebreak Lorentz-covariant perturbation theory is
investigated. A method for finding the difference between
separate terms of the corresponding series without their
explicit evaluation is proposed. A procedure of constructing
additional counter-terms to the canonical Hamiltonian that
compensate this difference at any finite order is proposed. For
the Yukawa model, the light-front Hamiltonian with all of these
counter-terms is obtained in a closed form. Possible application
of this approach to gauge theories is discussed.

\end{abstract}

\vfill
\noindent
Published in Theoretical and Mathematical Physics,
Vol.~112, No.~3, 1997.

\noindent
Translated from Teoreticheskaya i Maternaticheskaya Fizika.
Vol.~112, No.~3,\\  pp.~399-416, September, 1997.

\newpage
\sect{vvede}{Introduction}

The Hamiltonian approach to quantum field theory  in  light-front
coordinates (LF),\linebreak
\hbox{$x_{\pm}={1\over\sqrt{2}}(x_0\pm x_3)$},
\hbox{$x^{\p}=\{x^1,x^2\}$},
is attractive as a  possible  method  of  solving
strong  interaction  problems.  In  this  approach,  the   formal
triviality of the physical vacuum allows one to seek bound states
without prior investigation  of  the  complex  vacuum  structure.
However, as is already known, canonical quantization in LF, i.e.,
on the $x^+=const$ hypersurface, can result in a theory not  quite
equivalent to the Lorentz-invariant theory (i.e., to the standard
Feynman  formalism).  This  is  due,  first  of  all,  to  strong
singularities  at  zero  values  of  the  "light-like"   momentum
variables \hbox{$Q_-={1\over\sqrt{2}}(Q_0-  Q_3)$}.
To restore the  equivalence  with  a
Lorentz-covariant theory, one has to add unusual counter-terms to
the formal canonical Hamiltonian for the LF,
$H=P_+={1\over\sqrt{2}}(P_0+P_3)$
(the operator of a  shift  along  the  $x^+$-axis).  These
counter-terms can be found by comparing the  perturbation  theory
based  on  the  canonical  LF  formalism  with  Lorentz-covariant
perturbation theory. This is  done  in  the  present  paper.  The
light-front  Hamiltonian  thus  obtained  can  then  be  used  in
nonperturbative  calculations.  It  is  possible,  however,  that
perturbation  theory  does  not  provide  all  of  the  necessary
additions  to  the  canonical  Hamiltonian,  as  some  of   these
additions can be nonperturbative. In spite  of  this,  it  seems
necessary  to  examine  this  problem  within  the  framework  of
perturbation theory first.

For practical purposes a stationary noncovariant
light-front perturbation theory,
which is similar to  the  one  applied  in
nonrelativistic quantum mechanics, is widely used. It  was  found
\cite{bur1,har,lay}
that the "light-front" Dyson formalism allows  this  theory
to be transformed into an equivalent light-front Feynman theory
(under an appropriate regularization). Then,  by  re-summing  the
integrands of the Feynman integrals, one can recast their form so
that they become the same as  in  the  Lorentz-covariant  theory.
(This is not the case for diagrams without external lines,  which
we do not  consider  here.)  Then,  the  difference  between  the
light-front and Lorentz-covariant  approaches  that  persists  is
only due to the different regularizations and  different  methods
of calculating the Feynman integrals (which is important  because
of  the  possible  absence  of  their  absolute  convergence   in
pseudo-Euclidean space). In the present paper, we concentrate  on
the analysis of this difference.

A  light-front  theory  needs  not  only    the    standard    UV
regularization,  but  also  a  special  regularization  of   the
singularities $Q_-=0$. In  our  approach,  this  regularization
eliminates  the  creation  operators  $a^+(Q)$  and    annihilation
operators $a(Q)$ with $|Q_-^i|<\e$ from the Fourier expansion  of  the
field operators in the
field representation. As a result,  the  integration  w.r.t.  the
corresponding momentum $Q_-$ over the range $(-\infty,-\e)\cup(\e,\infty)$
is associated  with  each  line  before  removing  the  $\d$-functions.
Different propagators are regularized independently, which allows
the described re-arrangement of the perturbation  theory  series.
On the other hand, this regularization is convenient for  further
nonperturbative  numerical  calculations  with  the   light-front
Hamiltonian, to which the necessary counter-terms are added  (the
"effective"  Hamiltonian).  We  require  that  this   Hamiltonian
generate a theory equivalent to the Lorentz-covariant theory when
the  regularization  is  removed.  Note  that   Lorentz-invariant
methods of regularization (e.g., Pauli-Villars regularization) are
far less convenient for numerical calculations and we shall  only
briefly mention them.

The specific properties  of  the  light-front  Feynman  formalism
manifest themselves only in the integration over the variables
$Q_{\pm}={1\over\sqrt{2}}(Q_0\pm  Q_3)$,
 while integration over the transverse momenta $Q_{\p}\equiv \{Q_1, Q_2\}$
is  the  same  in  the  light-front  and  the    Lorentz
coordinates (though it might be nontrivial  because  it  requires
regularization and renormalization). Therefore, we concentrate on
a comparison of diagrams for fixed transverse momenta  (which  is
equivalent to a two-dimensional problem).

In the present paper, we propose a method that allows one to find
the difference (in the limit $\e\to 0$)  between  any  light-front
Feynman integral and the corresponding Lorentz-covariant integral
without having  to  calculate  them  completely.  Based  on  this
method, a procedure is elaborated for constructing  an  effective
Hamiltonian in LF  in  any  order  of  perturbation  theory.  The
procedure can be applied to all nongauge field theories, as  well
as to Abelian and non-Abelian gauge theories in the gauge $A_-=0$
with the vector  meson  propagator  chosen  according  to  the
Mandelstam-Leibbrandt prescription \cite{man,lei}. The question of whether  the
additional components  of  the  Hamiltonian  that  arise  can  be
combined into a finite number of counter-terms must be dealt with
separately in each particular case.

Application of this  formalism  to  the  Yukawa  model  makes  it
possible to obtain the effective  light-front  Hamiltonian  in  a
closed form. The result agrees with the conclusions of
the work  \cite{bur1}, where
a comparison was made of the  light-front  and  Lorentz-covariant
methods via calculating self-energy  diagrams  in  all  orders  of
perturbation  theory  and  other  diagrams  in   lowest    orders.
Conversely, for gauge theories (both Abelian and non-Abelian), it
was found that counter-terms of arbitrarily high order  in  field
operators must be added to the effective Hamiltonian. This result
may  turn  out  to  be  wrong  if  the  contributions   to    the
counter-terms are  mutually  canceled.  This  calls  for  further
investigation, but such possibility  appears to be very unlikely.

What we have said  above  does  not  depreciate  the  light-front
formalism as applied to gauge theories. This is because the  only
requirement concerning the light-front Hamiltonian is that  it  correctly
reproduces all gauge-invariant  quantities  rather  than  the
off-mass-shell Feynman  integrals  in  a  given  gauge.  However,
renormalization of the light-front Hamiltonian turns out to be  a
difficult problem and it requires  new  approaches.  We  do  not
examine the possibilities of changing the light-front Hamiltonian
by introducing new nonphysical fields by a method different  from
the Pauli-Villars regularization  \cite{sold}  or  the  possibilities  of
using  gauges  more  general  than   $A_-=0$  with    the
Mandelstam-Leibbrandt propagator. These points also  need  to  be
investigated further.

\sect{integ}{Reduction of light-front and Lorentz-covariant \st
Feynman integrals to a form convenient for comparison}

Let us examine an arbitrary IPI Feynman diagram. We fix all external momenta and all transverse
momenta of integration, and integrate only over
$Q_+$ and $Q_-$:
 \disn{1}{
F=\lim_{\k\to 0}\int{{\prod_i d^2Q^i \quad  f(Q^i,p^k)}
\over  {\prod_i(2Q_+^iQ_-^i-M^2_i+i\k)}}.
\nom}
We assume that all vertices are polynomial and that the propagator has the form
 \disn{1.2}{
{{z(Q)}\over {Q^2-m^2+i\k}},\qquad{\rm or}\qquad {{z(Q)\; Q_+}
\over{(Q^2-m^2+i\k)(2Q_+Q_-+i\k)}},
\nom}
where $z(Q)$ is a polynomial.
A propagator of the second type in (2) arises in gauge theories
in the gauge  $A_-=0$
if the Mandelstam-Leibbrandt formalism \cite{man,lei}
with the vector boson propagator
 \dis{
\frac{1}{Q^2+i\k}\( g_{\mu\nu}-\frac{(\d_\mu^+Q_\nu+Q_\mu \d_\nu^+)2Q_+}
{2Q_+Q_-+i\k}\),
}
is used.
In
Eq.~(\ref{1}) either $M^2_i=m^2_i+{Q^{i}_{\p}}^2\ne 0$, where  $m_i$
is the particle mass, or $M^2_i=0$.

The function  $f$
involves the numerators of all propagators and all vertices
with the necessary
$\d$-functions,
that include the external momenta  $p^k$  (the same expression without the
$\d$-functions is a polynomial, which we denote by $\tilde  f$).
We assume for the diagram  $F$  and for all of its subdiagrams that
the conditions
 \disn{1.1}{
\o_{\pa}<0, \qquad \o_+<0,
\nom}
hold, where $\o_+$ is the index of divergence w.r.t.
 $Q_+$ at $Q_-^i\ne 0\  \forall  i$,  and  $\o_{\pa}$
is the index of divergence in $Q_+$ and $Q_-$ (simultaneously);
 \hbox{$Q_{\pm}={1\over\sqrt{2}}(Q_0\pm  Q_3)$}.
The diagrams that do not meet these conditions should
be examined separately for each particular theory
(their number is usually finite). We
seek the difference
between the value of integral
(\ref{1})
obtained by the Lorentz-covariant calculation and its value calculated in
light-front coordinates (light-front calculation).

In the light-front calculation, one introduces and
then removes the light-front cutoff
$|Q_-|\ge\e>0$:
 \dis{
F_{\rm lf}=\lim_{\e\to 0}\lim_{\k\to 0}
\int\limits_{V_{\scriptstyle \e}}\prod_i dQ_-^i \int
\prod_i dQ_+^i
{{f(Q^i,p^k)}\over{\prod_i (2Q_+^iQ_-^i-M^2_i+i\k)}},
}
where
 $V_{\e}=\prod_i\(\(-\infty,-\e\)\cup\(\e,\infty\)\)$.
Here (and in the diagram configurations to be defined below) we take
the limit w.r.t. $\e$,
but, generally speaking, this limit may not exist. In this case, we assume
that we do not
take the limit, but take the sum of all nonpositive power terms of the
Laurent series in $\e$ at the zero point.
If conditions
(\ref{1.1}) are satisfied, Statement 2 from Appendix I can be used.
This results in the equality
 \disn{5}{
F_{\rm lf}=\lim_{\e\to 0}\lim_{\k\to 0}\int\prod_k dq_+^k
\int\limits_{V_{\scriptstyle \e}\cap B_L} \prod_k dq_-^k
{{\tilde f(Q^i,p^s)}
\over{\prod_i (2Q_+^iQ_-^i-M^2_i+i\k)}}.
\nom}
From here on, the momenta of the lines $Q^i$
are assumed to be expressed in terms of the loop momenta
$q^k$,  $B_L$  is a sphere of a radius  $L$ in the $q_-^k$-space,  and
$L$ depends on the external momenta. Now, using Statement 2 from
Appendix I, we obtain
 \disn{7}{
F_{\rm lf}=\lim_{\e\to 0}\lim_{\k\to 0}
\lim_{\b\to 0}\lim_{\g\to 0}
\int\prod_k dq_+^k \int\limits_{V_{\scriptstyle \e}}
\prod_k dq_-^k
{{\tilde f(Q^i,p^s) \; e^{-\g\sum_i {Q_+^i}^2-\b \sum_i{Q_-^i}^2}
}\over{\prod_i (2Q_+^iQ_-^i-M^2_i+i\k)}}.
\nom}
To reduce the covariant Feynman integral to a form similar to
(\ref{5}), we introduce a quantity $\hat F$:
 \disn{8}{
\hat F=\lim_{\k\to  0}\lim_{\b\to 0}\lim_{\g\to 0}
\int  \prod_k  d^2q^k  {{\tilde  f(Q^i,p^s)  \;
e^{-\g\sum_i    {Q_+^i}^2-\b
\sum_i{Q_-^i}^2}}\over{\prod_i (2Q_+^iQ_-^i-M^2_i+i\k)}}.
\nom}
Let us prove that this quantity coincides with the result of the
Lorentz-covariant calculation $F_{\rm  cov}$. To this end, we introduce the
\hbox{$\a$-representation}  in the Minkowski space of the propagator
 \disn{9}{
{{z(Q^i)}\over{2Q_+^iQ_-^i-M^2_i+i\k}}
=-iz\(-i{{\dd}\over{\dd y_i}}\)\int\limits_0^{\infty}
e^{i\a_i(2Q_+^iQ_-^i-M^2_i+i\k)+i(Q^i_+y_i^++Q^i_-y_i^-)}
d\a_i \Bigr|_{y_i=0}.
\nom}
Then we substitute (\ref{9}) into (\ref{8}). Due to the exponentials
that cut off $q_+^k$, $q_-^k$ and $\a^i$ the integral over these
variables is absolutely convergent. Therefore, one can interchange
the integrations over $q_+^k$,
$q_-^k$ and $\a^i$. As
a result, we obtain the equality
 \disn{18}{
\hat F=\lim_{\k\to 0}\lim_{\b\to 0}\lim_{\g\to 0}
\int\limits_0^\infty \prod_n d\a_i \;\hat \f(\a_i,p^s,\g,\b)
\; e^{-\k\sum_i\a_i},
\nom}
where
 \disn{19}{
\hat \f(\a_i,p^s,\g,\b)=(-i)^n
\tilde f\(-i{{\dd}\over{\dd y_i}}\)\times \no
\times \int \prod_k d^2q^k \;
e^{\sum_i \[i\a_i(2Q_+^iQ_-^i-M^2_i)+
i(Q^i_+y_i^++Q^i_-y_i^-)-\g{Q_+^i}^2-\b{Q_-^i}^2\]}
\Bigr|_{y_i=0}.
\nom}
For the Lorentz-covariant calculation in the \hbox{$\a$-representation}
satisfying conditions (\ref{1.1}), there is a known
expression \cite{bo}
 \disn{17}{
F_{\rm cov}=\lim_{\k\to 0}
\int\limits_0^\infty \prod_n d\a_i \;\f_{\rm cov}(\a_i,p^s)
\; e^{-\k\sum_i\a_i},
\nom}
where
 \disn{16}{
\f_{\rm cov}(\a_i,p^s)=(-i)^n \tilde f\(-i{{\dd}\over
{\dd y_i}}\)\times\no
\times \lim_{\g,\b\to 0} \int \prod_k d^2q^k
\; e^{\sum_i \[i\a_i(2Q_+^iQ_-^i-M^2_i)+
i(Q^i_+y_i^++Q^i_-y_i^-)-\g{Q_+^i}^2-\b{Q_-^i}^2\]}
\Bigr|_{y_i=0}.
\nom}
In Appendix 2, it is shown that in (\ref{18}) the limits in
$\g$ and $\b$ can be interchanged, in turn, with the integration
over
$\{\a_i\}$, and then with $\tilde f\(-i{{\dd}\over{\dd  y_i}}\)$.
After that, a comparison of relations (\ref{18}), (\ref{19}) and (\ref{17}),
(\ref{16}),  clearly shows that $\hat F=F_{\rm   cov}$.
Considering  (\ref{8})  and using Statement 1 from Appendix 1,
we obtain the equality
 \disn{20}{
F_{\rm cov}=\lim_{\k\to  0}\int\prod_k  dq_+^k
\int\limits_{B_L}
\prod_k    dq_-^k    {{\tilde    f(Q^i,p^s)}
\over{\prod_i (2Q_+^iQ_-^i-M^2_i+i\k)}}.
\nom}
Expression (\ref{20}) differs from  (\ref{5})
only by the range of the integration over $q_-^k$.

\sect{polos}{Reduction of the difference between the light-front
and\st  Lorentz-covariant Feynman integrals
to a sum of configurations}

Let us introduce a partition for each line,
 \disn{22}{
\( \int\limits_{-\infty}^{-\e}
dQ_- + \int\limits_{\e}^{\infty} dQ_- \)=
\[\int dQ_-+(-1)\int\limits_{-\e}^{\e} dQ_-\].
\nom}
We call a line with integration w.r.t. the momentum  $Q_-^i$  in  the
range $(-\e,\e)$ (before removing $\d$-functions) a type-1 line, a
line with integration in the range $(-\infty,-\e)\cup(\e,\infty)$
a  type-2
line, and a line with integration over the whole range
$(-\infty,\infty)$
a full line. In the  diagrams,  they  are  denoted  as  shown  in
Figs.~la, b, and c, respectively.
 \begin{figure}[ht]
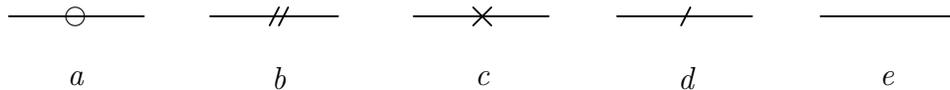

\input fig1.pic
\caption{Notation for different types of lines  in  the  diagrams:
"a" is a type-1 line, "b" is a type-2 line, "c" is a  full  line,
"d" is an $\e$-line, and "e" is a $\Pi$-line.}
\end{figure}

Let us substitute partition (\ref{22}) into expression (\ref{5})
for $F_{\rm  lf}$ and
open the brackets. Among the resulting terms,  there  is  $F_{\rm cov}$
(expression  (\ref{20})).  We  call  the  remaining   terms    "diagram
configurations" and denote them by $F_j$. Then  we  arrive  at  the
relation  $F_{\rm lf}-F_{\rm cov}=\sum\limits_jF_j$, where
 \disn{24}{
F_j=\lim_{\e\to 0}\lim_{\k\to 0}\int\prod_k dq_+^k
\int\limits_{V^j_{\scriptstyle \e}\cap B_L} \prod_k dq_-^k
{{\tilde f(Q^i,p^s)}
\over{\prod_i (2Q_+^iQ_-^i-M^2_i+i\k)}},
\nom}
and $V^j_{\e}$ is the region corresponding to  the  arrangement  of  full
lines and type-1 lines in the given configuration.

Note that before taking the limit in $\e$, Eqs. (\ref{20}) and  (\ref{24})
can be used successfully: first, they are applied to a  subdiagram  and,
then, are substituted into the formula for  the  entire  diagram.
This is admissible because, after the deformation of the contours
described in the proof  of  Statement  1  from  Appendix  1,  the
integral over the loop momenta $\{q_+^k\}$ of the  subdiagram  converges
(after integration over the variables $\{q_-^k\}$  of  this  subdiagram)
absolutely and uniformly  with  respect  to  the  remaining  loop
momenta $\{q_-^{k'}\}$. Therefore, one can interchange the  integrals  over
$\{q_+^k\}$ and $\{q_-^{k'}\}$.

Thus,   the    difference    between    the    light-front    and
Lorentz-covariant calculations of the diagram is given by the sum
of all of its configurations. A configuration of a diagram is the
same diagram, but where each line is labeled as a full or  type-1
line, provided that at least one type-1 line exists.

\sect{epsil}{Behavior of the configuration as  $\e\to 0$}

We  assume  that  all
external momenta $p^s$ are fixed for the diagram in question and
 \disn{D0}{
p_-^s\ne 0, \quad \sum_{s'}p_-^{s'}\ne 0,
\nom}
where the summation is taken over any subset  of  external  momenta;
all of these momenta are assumed to be directed inward.

Let us consider an arbitrary configuration.  We  apply  the  term
"$\e$-line" to all type-1 lines  and  those  full  lines  for  which
integration over $Q_-$ actually does not  expand  outside  the  domain
$(-r\e,r\e)$, where  $r$ is a finite number  (below,  we  explain  when
these lines appear). The remaining full lines are called $\Pi$-lines.
In the diagrams, the $\e$-lines and $\Pi$-lines are denoted as shown  in
Figs.~1d and e, respectively. Note that the diagram can be  drawn
with  lines  "a"  and  "c"  from  Fig.~1  (this   defines    the
configuration unambiguously), or with lines "d" and "e" (then the
configuration is not uniquely defined).

If among the lines arriving at the vertex only one is  full
and the others are type-1 lines, this full line is an  $\e$-line  by
virtue of the momentum conservation at the vertex. The  remaining
full lines form a subdiagram (probably unconnected). By virtue of
conditions (\ref{D0}), there is a connected part to which  all  of  the
external lines are attached. All of the  external  lines  of  the
remaining  connected  parts  are   $\e$-lines.  Consequently,   using
Statement~1 from Appendix~1, we can see that integration over the
internal momenta of these connected parts can be carried out in a
domain of order $\e$ in size, i.e., all of their internal lines  are
$\e$-lines. Thus, an arbitrary configuration can be drawn as in Fig.~2
and integral (\ref{24}), with the corresponding  integration  domain,
is associated with it.
 \begin{figure}[ht]
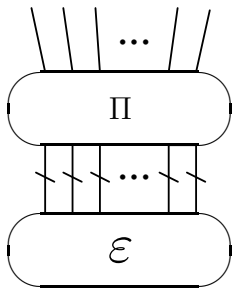

\input fig2.pic
\caption{
Form of an arbitrary configuration: $\Pi$ is  the  connected
subdiagram consisting of $\Pi$-lines, {\protect\LARGE $\e$}
is the subdiagram  consisting
of $\e$-lines and, probably, containing no vertices.}
\end{figure}

Let us investigate the behavior of the configuration  as  $\e\to 0$.
From here on, it is convenient to represent the propagator as
 \disn{D1}{
{{\tilde z(Q)}\over {Q^2-m^2+i\k}},\quad{\rm where}\quad
 \tilde z(Q)=z(Q)\; \; {\rm or}\; \; \tilde z(Q)={{z(Q)}\over
{2Q_-+i\k/Q_+}}.
\nom}
rather than as (\ref{1.2}). Then, in (\ref{1}),
$M^2_i=m^2_i+{Q^{i}_{\p}}^2\ne 0$  and  the
function $\tilde f$ is no longer a polynomial. If  the  numerator  of  the
integrand consists  of  several  terms,  we  consider  each  term
separately (except when  the  terms  arise  from  expressing  the
propagator momentum $Q_-^i$ in terms of loop and external momenta).

We denote the loop momenta of subdiagram $\Pi$ in Fig.~2 by  $q^l$ and
the others  by  $k^m$.  We  make  following  change  of  integration
variables in (\ref{24}):
 \disn{D10}{
k_-^m \to \e \; k_-^m.
\nom}
Then,  the  integration  over  $k_-^m$  goes  within  finite    limits
independent of $\e$. We denote the power of $\e$ in the  common  factor
by $\t$ (it stems from the volume elements  and  the  numerators
when the transformation (\ref{D10}) is made).
The contribution to $\t$ from the expression
$1/(2Q_-+i\k/Q_+)$
(Eq.~(\ref{D1})), which is related to the $\e$-line, is equal  to  -1.  We
divide the domain of integration over $k_+^m$ and $q_+^l$ into sectors such
that the momenta of all full lines $Q_+^i$ have the same  sign  within
one sector.

In Statement~1 of Appendix~1, it is shown that for  each  sector,
the contours of integration over $q_-^l$ and $k_-^m$ can be bent in such a
way that absolute convergence in $q_+^l$,  $k_+^m$,
$q_-^l$  and $k_-^m$ takes place.
Since, in this case, the momenta $Q_-^i$  of  $\Pi$-lines  are  separated
from  zero  by  an $\e$-independent  constant,  the   corresponding
$\Pi$-line-related propagators and factors from the vertices  can  be
expanded in a series in $\e$. This expansion commutes with integration.

It is also clear that the denominators of the  propagators  allow
the following estimates under an infinite increase in $|Q_+|$:
 \dis{\hfill
\left|{1\over{2Q_+ Q_--M^2+i\k}}\right|\le \cases{
{\displaystyle {1\over{c \; |Q_+|}}} & for $\Pi$-lines,
\refstepcounter{form} \label{D11} \hskip 40mm \hfill (\arabic{form})\cr
{\displaystyle {1\over{\tilde c\; \e \; |Q_+|}}} & for $\e$-lines,
\refstepcounter{form} \label{D12} \hfill (\arabic{form})\cr}
}
Here $c$ and $\tilde c$ are $\e$-independent constants.
Note  that  for  fixed
finite $Q_+$, the estimated expressions are bounded as $\e\to  0$. After
transformation (\ref{D10}) and release of the factor ${1\over{\e}}$
(in  accordance
with  what  was  said  about  the  contribution  to $\t$),    the
$\e$-line-related expression from (\ref{D1}) becomes
 \dis{
\left| {1\over{2Q_-+i\k/Q_+}}\right| \to
\left| {1\over{2Q_-+i\k/(Q_+\e)}}\right| \le
{1\over{2|Q_-^i|}},
}
where a $Q_+$-independent quantity was used for the  estimate  (this
quantity is meaningful and does not  depend  on  $\e$  because  the
value of $Q_-$ is separated from zero by an $\e$-independent constant).

We integrate first over $q_+^l$, $k_+^m$ within one sector  and  then  over
$q_-^l$,  $k_-^m$  (the latter integral converges uniformly in  $\e$).
Let  us
examine the  convergence  of  the  integral  over  $q_+^l$,  $k_+^m$   with
canceled denominators of the $\e$-lines (which is equivalent to
estimating expressions (\ref{D12}) by a constant). If it converges, then
the initial integral  is  obviously  independent  of  $\e$  and  the
contribution  from  this  sector  to   the    configuration    is
proportional to $\e^\t$.

Let us show that if it diverges with a degree of  divergence  $\a$,
the contribution to the initial integral is proportional to $\e^{\t-\a}$
up to logarithmic corrections. To this end, we divide the  domain
of integration over $q_+^l$,
$k_+^m$ into  two  regions:  $U_1$,  which  lies
inside a sphere of radius $\L/\e$ ($\L$ is fixed), and $U_2$,  which  lies
outside this sphere (recall that in our reasoning, we  deal  with
each sector separately). Now we estimate (\ref{D11})
(like  (\ref{D12}))  in
terms of ${\displaystyle{1\over{\hat  c\e|Q_+|}}}$
(which is admissible) and change  the  integration
variables as follows:
 \disn{D13}{
q_+^l\to {1\over{\e}}\; q_+^l,\quad k_+^m\to {1\over{\e}}\; k_+^m.
\nom}
After $\e$ is factored out of the numerator and the volume  element,
the integrand becomes  independent  of  $\e$.  Thus,  the  integral
converges.

One can choose such $\L$ (independent of  $\e$) that  the  contribution
from the  domain  $U_2$ is  smaller  in  absolute  value  than  the
contribution from the domain $U_1$. Consequently, the whole integral
can he estimated
via the integral over the  finite  domain  $U_1$.  Now  we  make  an
inverse replacement in (\ref{D13}) and estimate (\ref{D12})
by a  constant  (as
above). Since the size of the integration domain is $\L/\e$  and  the
degree of divergence is $\a$, the integral behaves as  $\e^{-\a}$  (up  to
logarithmic corrections), q.e.d. This reasoning is valid for each
sector and, thus, for the configuration as a whole. Obviously,
 \disn{D13.1}{
\a=\max\limits_{r}\a_r,
\nom}
where $\a_r$ is the subdiagram divergence index and the  maximum  is
taken over all subdiagrams $D_r$ (including unconnected  subdiagrams
for which $\a_r$ is the sum  of  the  divergence  indices  of  their
connected parts). In the case under consideration, $\a_r=\o_+^r+\n^r$,
where $\n^r$ is the number of internal $\e$-lines in the subdiagram
$D_r$.
The  quantities  $\o_{\pm}^r$  are  the  UV  divergence  indices  of   the
subdiagram  $D_r$ w.r.t. $Q_{\pm}$.

Above, we introduced a quantity $\t$, which is equal to the power of
$\e$ that stems from the  numerators  and  volume  elements  of  the
entire configuration. We can write $\t=\o_-^r-\m^r+\n^r+\eta^r$,  where
$\m^r$ is the index of the UV  divergence  in  $Q_-$ of a smaller subdiagram
(probably, a tree subdiagram or a nonconnected one) consisting of
$\Pi$-lines entering $D_r$. The term $\eta^r$
is the power of $\e$ in the  common
factor, which, during transformation (\ref{D10}), stems from the  volume
elements and numerators of the lines that did not enter $D_r$. (It is
implied that the integration momenta are chosen in the  same  way
as when calculating the divergence indices of $D_r$.)  Then,  up  to
logarithmic corrections, we have
 \disn{13.4}{
F_j\sim \e^{\s},\quad \s=\min_r(\t,\o_-^r-\o_+^r-\m^r+\eta^r).
\nom}
Consequently, for $\e\to 0$, the configuration is equal to zero if
$\s>0$. Relation (\ref{13.4})
allows  all  essential  configurations  to  be
distinguished.

\sect{ispra}{Correction procedure and analysis of counter-terms}

We want to build a corrected light-front Hamiltonian
$H_{\rm  lf}^{\rm  cor}$ with  the
cutoff $|Q_-^i|>\e$, which would generate  Green's  functions  that
coincide in the limit $\e  \to  0$  with  covariant  Green's  functions
within the perturbation theory. We begin with a  usual  canonical
Hamiltonian in the light-front coordinates  $H_{\rm lf}$  with  the  cutoff
$|Q_-^i|>\e$. We imply that the integrands of the  Feynman  diagrams
derived from  this  light-front  Hamiltonian  coincide  with  the
covariant integrands after some  resummation  \cite{bur1,har,lay}.
However,  a
difference may arise due to the  various  methods  of  doing  the
integration, e.g., due to different auxiliary regularizations. As
shown in Sec.~\ref{polos}, this difference (in the limit
$\e  \to  0$) is equal to
the sum of all properly arranged configurations of  the  diagram.
One should  add  such  correcting  counter-terms  to  $H_{\rm lf}$,  which
generates  additional  "counter-term"  diagrams,  that  reproduce
nonzero (after taking limit w.r.t. $\e$) configurations  of  all  of
the diagrams. Were we able  to  do  this,  we  would  obtain  the
desired $H_{\rm lf}^{\rm cor}$.
In fact, we can only show how to  seek  the  $H_{\rm lf}^{\rm cor}$  that
generates the Green's functions  coinciding  with  the  covariant
ones everywhere except the null  set  in  the  external  momentum
space (defined by condition (\ref{D0})). However, this  restriction  is
not essential because this possible difference  does  not  affect
the physical results.

Our  correction  procedure  is  similar  to  the  renormalization
procedure. We assume that the perturbation  theory  parameter  is
the number of loops. We carry out the correction by steps: first,
we find the counterterms to the  Hamiltonian  that  generate  all
nonzero configurations of the diagrams up to the given order and,
then, pass to the next order. We take into account that this step
involves  the  counter-term  diagrams  that  arose    from    the
counter-terms added to the Hamiltonian for lower orders. Thus, at
each  step,  we  introduce  new  correcting  counter-terms   that
generate the difference remaining in this order. Let us show  how
to successfully look for the correcting counter-terms.

We call a configuration nonzero if it does not vanish as $\e\to 0$.
We  call  a  nonzero  configuration  "primary"  if $\Pi$  is  a   tree
subdiagram in it (see Fig.~2). Note that for this  configuration,
breaking any  $\Pi$-line results in a violation of  conditions  (\ref{D0});
then, the resulting diagram is not a configuration. We  say  that
the configuration is changed if all of the $\Pi$-lines in the related
integral (\ref{24})  are expanded in series  in  $\e$  (see  the  reasoning
above Eq.~(\ref{D11}) in Sec.~\ref{epsil}) and only those terms that
do not vanish
in the limit $\e\to 0$ after the
integration are retained. As  mentioned  above,  developing  this
series and integration are interchangeable operations.  Thus,  in
the limit $\e  \to  0$,  the  changed  and  unchanged  configurations
coincide. Therefore,  we  always  require  that  the  Hamiltonian
counter-terms generate changed configurations, as this simplifies
the form of the counter-terms.  Using  additional  terms  in  the
Hamiltonian, one can generate only counter-term  diagrams,  which
are equal to zero for  external  momenta  meeting  the  condition
$|p_-^s|<\e$, because with the cutoff used (see  the  Introduction),
the external lines of the diagrams  do  not  carry  momenta  with
$|p_-^s|<\e$. We bear this in mind in what follows.

We seek counter-terms by the induction method. It is clear  that,
in  the  first  order  in  the  number  of  loops,  all   nonzero
configurations  are  primary.  We  add  the  counter-terms   that
generate them to the Hamiltonian. Now, we  examine  an  arbitrary
order of perturbation theory. We assume that in lower orders, all
nonzero  configurations  that  can   be    derived    from    the
counter-terms, accounting for the  above  comment,  have  already
been generated by the Hamiltonian.

Let us proceed to  the  order  in  question.  First,  we  examine
nonzero configurations with only one loop momentum $k$ and a number
of momenta $q$ (see the notation above  Eq.~(\ref{D10})).  We  break  the
configuration lines one by one without touching the  other  lines
(so that the ends of the broken lines become external lines). The
line break may result in a structure that is not a  configuration
(if conditions (\ref{D0}) are violated); a line break may  also  result
in a zero configuration or in a  nonzero  configuration.  If  the
first case is realized for each broken line, then  the  initial
configuration  is  primary  and  it  must  be  generated  by  the
counter-terms  of  the  Hamiltonian   in    the    order    under
consideration. If breaking of each line results in either  the  first
or the second case, we call the initial configuration real and it
must be also generated in this order.

Assume that breaking a line results in the third case. This means
that the resulting configuration stems from counter-terms in the
lower orders. Then, after restoration of the broken  line  (i.e.,
after  the  appropriate  integration),  it  turns  out  that  the
counter-terms of the lower  orders  have  generated  the  initial
configuration (we take into account  the  comment  on  successive
application of Eq.~(\ref{24});  see  the  end  of
Sec.~\ref{polos})  with  the
following distinctions: (i) the broken line (and, probably,  some
others, if a nonsimply connected diagram  arises  after  breaking
the line) is  not  a  $\Pi$-line  but  a  type-2  line,  due  to  the
conditions $|p_-^s|>\e$; (ii) if, after restoration  of  the  broken
line, the behavior at small $\e$ becomes worse (i.e., $\s$ decreased),
then fewer terms than are necessary for the initial configuration
were considered in the above-mentioned series  in  $\e$.  We  expand
these arising type-2 lines by formula  (\ref{22})  and  obtain  a  term
where all of these lines are replaced by $\Pi$-lines or  other  terms
where some (or all) of these lines have become type-1  lines.  In
the latter case, one of the momenta $q$ becomes the momentum $k$.  We
call these  terms  "repeated  parts  of  the  configuration"  and
analyze them together  with  the  configurations  that  have  two
momenta  $k$.  In  the  former  case,  we  obtain    the    initial
configuration up to distinction (ii). We add  a  counter-term  to
the  Hamiltonian  that  compensates  this  distinction    (the
counter-term diagrams generated by it are called the compensating
diagrams).

If there is only one line for which the third case  is  realized,
it turns out that, in the given order, it is  not  necessary  to
generate the initial configuration by the  counter-terms,  except
for the compensating addition  and  the  repeated  part  that  is
considered at the next step. If there are several lines for which
the  third  case  is  realized,  the  initial  configuration   is
generated in lower orders more than once.  For  compensation,  it
should be generated (with the corresponding numerical coefficient
and the opposite sign) by the Hamiltonian  counter-terms  in  the
given order. We call this configuration a secondary one. Next, we
proceed to examine configurations with two momenta $k$ and so on up
to  configurations  with  all  momenta  $k$,  which  are   primary
configurations.

Thus, the configurations  to  be  generated  by  the  Hamiltonian
counter-terms can be  primary  (not  only  the  initial  primary
configurations but also the repeated parts  analogous  to  them,
called primary-like),
real, compensating, and  secondary.  If  the  theory  does  not
produce either the loop consisting only of lines with $Q_+$  in  the
numerator (accounting for contributions from the vertices)  or  a
line with ${Q_+}^n$ in the numerator for $n>1$, then real configurations
are absent because a line without $Q_+$ in the numerator can  always
be broken without  increasing $\s$  (see  Eq.~(\ref{13.4})).  It  is  not
difficult to demonstrate that if  each appearing  primary, real,  and
compensating configuration has only two external line,
then there are  no  secondary configurations  at  all.

The  dependence    of    the    primary
configuration on external momenta becomes trivial if its degree of
divergence $\a$ is positive, the maximum in formula (\ref{D13.1})
is  reached
on the diagram itself, and $\s=0$. Then, only the first  term  is
taken into account in the above-mentioned series. Thus,  not  all
of the $\Pi$-line-related propagators and vertex  factors  depend  on
$k_-^m$ and they can be pulled out of the sign of the integral w.r.t.
$\{k_-^m\}$ in (\ref{24}). We then obtain
 \disn{D14}{
F_j^{\rm prim}=\lim_{\e\to  0}\lim_{\k\to 0}\int\prod_m dk_+^m
{{\tilde f'(k^m,p^s)}\over{\prod_i (2Q_+^iQ_-^i-M^2_i+i\k)}} \times \no
\times \int\limits_{V_{\scriptstyle \e}}  \prod_m   dk_-^m
{{\tilde f''(k^m)} \over{\prod_k (2Q_+^kQ_-^k-M^2_k+i\k)}},
\nom}
where $V_{\e}$ is a domain  of  order  $\e$  in  size.  Let  us  carry  out
transformations (\ref{D10}) and (\ref{D13}).
For the denominator of the $\Pi$-line,
we obtain
 \dis{
{1\over{2({1\over{\e}}\sum k_+ +\sum p_+)(\sum p_-)-M^2+i\k}}\to
{{\e}\over{2(\sum k_+)(\sum p_-)}}.
}
Here we neglect terms of order $\e$ in the denominator  because  the
singularity at $k_+^m=0$ is integrable under the  given  conditions
for $\a$ and everything can be calculated in zero order in $\e$ at $\s=0$.
Thus, the dependence on external  momenta  can  be  completely
collected into an easily obtained common factor.

\sect{jukav}{Application to the Yukawa model}

The Yukawa model involves diagrams that do not satisfy  condition
(\ref{1.1}). These are displayed in Figs.~3a~and~b. We have
$\o_{\pa}=0$  for
diagram "a" and $\o_+=0$ for diagram "b".
 \begin{figure}[ht]
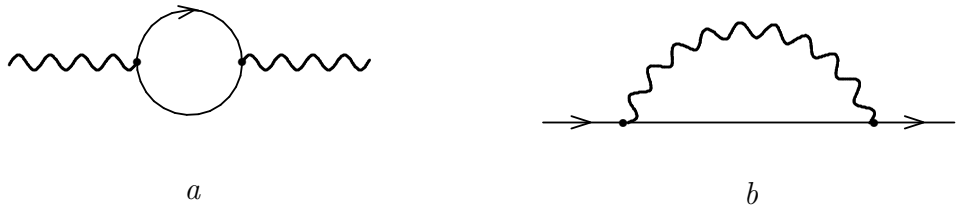

\input fig3.pic
\caption{Yukawa model diagrams that do not meet condition
(\protect\ref{1.1}).}
\end{figure}

Nevertheless, these  diagrams  can  be  easily  included  in  the
general scheme of reasoning. To this end, one should subtract the
divergent part, independent of external momenta, in the integrand
of the logarithmically divergent (in two-dimensional space,  with
fixed internal transverse momenta)  diagram  "a".  We  obtain  an
expression with $\o_{\pa}<0$ (i.e., which converges  in  two-dimensional
space) and $\o_+=0$, as  in  diagram  "b".  This  means  that  the
integral over $q_+$ converges only in the  sense  of  the  principal
value (and it is this value of the integral that should be  taken
in the light-front  coordinates  to  ensure  agreement  with  the
stationary noncovariant perturbation theory). This value  can  be
obtained by distinguishing the $q_+$-even part of the integrand.

Two approaches are possible. One is to introduce  an  appropriate
regularization in transverse momenta  and  to  imply  integration
over them; then, it is convenient to distinguish the part that is
even in four-dimensional momenta $q$.  The  other  is  to  keep  all
transverse  momenta  fixed;  then,  the  part  that  is  even  in
longitudinal momenta $q_{\pa}$ can be released. For the  Yukawa  theory,
we use the first approach. For the transverse regularization,  we
use a "smearing" of vertices, which  is  equivalent  to  dividing
each propagator by $1+{Q_{\p}^i}^2/{\L_{\p}}^2$.
In  four-dimensional  space,
diagram  "a"  diverges  quadratically.  Under  introduction   and
subsequent  removal  of  the  transverse   regularization,    the
divergent  part,  which  was  previously  subtracted  from   this
diagram, acquires the form $C_1+C_2\> p_{\p}^2$.

After separating the even part of the regularized expression,  we
fix all of the transverse momenta again. Then it turns  out  that
diagrams "a" and "b" in Fig.~3 meet conditions (\ref{1.1})  and  one  can
show that after all of the operations mentioned, the exponent  $\s$
(see (\ref{13.4})) does not decrease for  any  of  their  configurations.
Hence, they can be included in the  general  scheme  without  any
additional corrections.

Let  us  first  analyze  the  primary  configurations  (see   the
definition in Sec.~\ref{ispra}). In the numerators, $k_-$  appears  only  in
the zero or one power and there are no loops where the numerators
of all of the lines contain $k_-$. Consequently, one always has
$\t>0$,  $\m^r\le  0$, and $\eta^r\ge 0$ (see the definitions in
Sec.~\ref{epsil}).  Analyzing the
properties of the expression $\o_-^r-\o_+^r$ for the Yukawa model diagrams,
we conclude from (\ref{13.4}) that $\s\ge  0$ always holds. The general  form
of the nonzero primary configurations with $\s=0$ is depicted in
Fig.~4. Note that they are all configurations with two external line.
 \begin{figure}[ht]
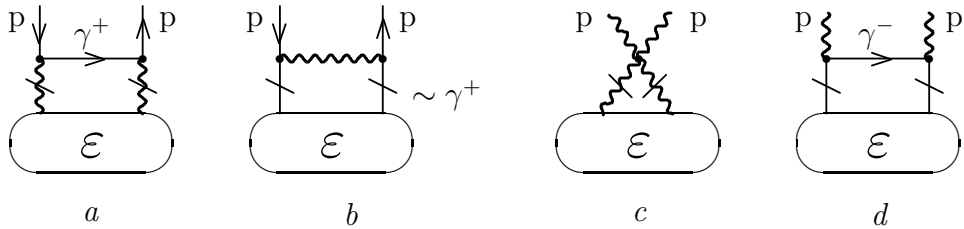

\input fig4.pic
\caption{
Nonzero configurations in the Yukawa model: $p$ is the external
momentum, and $\g^+$ or $\g^-$
symbols on the line indicate that the corresponding term is taken
in the numerator of the propagator.
In configuration "b", the part that is proportional to $\g^+$ is taken.}
\end{figure}

Further,  it  is  clear  that  there  are   no    nonzero    real
configurations (see the comment at the end of Sec.~\ref{ispra}), and it can
be shown by induction that there are no nonzero  compensating  or
secondary configurations either (the  definitions  are  given  in
Sec.~\ref{ispra} also). Thus, only primary or  primary-like  configurations
can be nonzero and all of them have the form shown in Fig.~4.  It
can be shown that their degree of divergence $\a$ is  positive  and
the maximum in formula (\ref{D13.1}) is reached for  the  diagram  itself.
Thus, the reasoning above and below formula (\ref{D14}) applies to them.
Then, denoting the configurations  displayed  in  Figs.~4a-d  by
\hbox{$D_{a}$~--~$D_{d}$}, we
arrive at the equalities
\hbox{$D_a={{\g^+}\over{p_-}}C_a$},
\hbox{$D_b={{\g^+}\over{p_-}}C_b$},             \hbox{$D_c=C_c$}
and \hbox{$D_d=C_d$},
where the expressions \hbox{$C_a$ -- $C_d$} depend only on the  masses  and
transverse momenta, but not on the external longitudinal momenta,
and have a finite limit as $\e\to 0$.

Now we assume that $D_{a}$  --  $D_{d}$
are not single  configurations  but  are
the sums  of  all  configurations  of  the  same  form  and  that
integration over the internal transverse momenta has already been
carried  out,  (with  the  above-described  regularization).   In
four-dimensional space, the diagrams $D_{a}$ and $D_{b}$ diverge  linearly
while $D_{c}$ and $D_{d}$ diverge quadratically. Therefore, because of  the
transverse regularization, the coefficients  $C_c$  and  $C_d$  in  the
limit for removing this regularization take the form $C_1+C_2\>  p_{\p}^2$,
where $C_1$ and $C_2$ do not depend on the external momenta (neither do
$C_a$, $C_b$)). Thus, to generate all  nonzero  configurations  by  the
light-front Hamiltonian, only the expression
 \disn{U1}{
H_c=\tilde C_1\; \f^2+\tilde C_2\> p_{\p}^2\; \f^2+
\tilde C_3\; \bar \psi\; {{\g^+}\over{p_-}}\; \psi,
\nom}
should be added, where $\f$ and $\psi$ are the boson and fermion fields,
respectively, and $\tilde C_i$, are the constant coefficients.

Comparing  (\ref{U1}) with   the    initial    canonical    light-front
Hamiltonian, one can easily see that the found counter-terms  are
reduced to a renormalization of various terms of the  Hamiltonian
(in particular the boson mass squared  and  the  fermion  mass
squared without changing the fermion mass itself).  The  explicit
Lorentz invariance is absent, which compensates the violation  of
the Lorentz invariance inherent, in the light-front formalism.

Note that in the framework of the second approach,  mentioned  at
the beginning of this section, one can obtain the  same  results.
The  only  difference  is  that  in  two-dimensional  space,  the
contributions from the configurations displayed in Fig.~3  would
additionally depend on external transverse momenta. However, this
dependence disappears after integration over internal  transverse
momenta with  the  introduction  and  subsequent  removal  of  an
appropriate regularization.

In the Pauli Villars regularization, it is easy  to  verify  that
the expression $\o_-^r-\o_+^r-\m^r+\eta^r$ from (\ref{13.4})
increases. This is
because the number of terms in the numerators of  the  propagator
increases. Then, the  contribution  from  the  $\e$-lines  does  not
change, while the $\Pi$-lines belonging to $D_r$ make  zero  contribution
to $\o_-^r-\o_+^r$ and $\eta^r$, but $-1$ contribution to $\m^r$.
Since $\t>0$, this
regularization makes it possible to meet
the condition $\s>0$ for the  configurations  that  were  nonzero
(one additional boson field and one additional fermion field  are
enough).  Then  it  turns  out  that  the  canonical  light-front
Hamiltonian cannot be corrected at all.

\sect{kalib}{Application to gauge theories}

Let us consider a gauge theory (e.g., QED or QCD) in the gauge
$A_-=0$.  The  boson  propagator  in    the    Mandelstam-Leibbrandt
prescription has the form
 \dis{
{1\over{Q^2+i\k}}\(g_{\m\n}-
{{Q_{\m}\d^+_{\n}Q_++Q_{\n}\d^+_{\m}Q_+}\over{2Q_+Q_-+i\k}}\).
}
All of the above reasoning was organized such that  it  could  be
applied to a theory like this (with fixed transverse momenta
$Q_{\p}\ne    0$). It turns out that there  are  nonzero  configurations  with
arbitrarily large numbers of external lines. An example of such a
configuration is given in Fig.~5.
 \begin{figure}[ht]
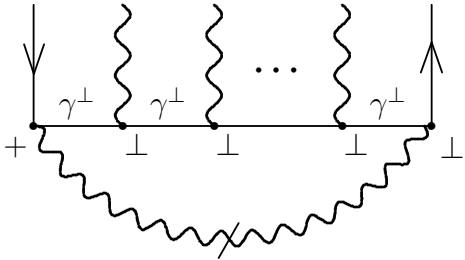

\input fig5.pic
\caption{
Nonzero configuration with an arbitrarily large number of external
lines in a gauge theory.
The symbols $\g^{\p}$ on the lines and the symbols $+$ or $\p$ by
the vertices indicate that the corresponding
terms $\g^+$ or $\g^+$ are taken in the numerators of
propagators and in the vertex factors.}
\end{figure}

Indeed, using formula (\ref{13.4}), we can see that for the configuration
in Fig.~5, $\t=0$ and, thus, $\s\le  0$,  i.e.,  this  is  a  nonzero
configuration.  It  is  also  clear  that  introduction  of   the
Pauli-Villars  regularization  does  not  improve  the  situation
because it does not affect $\t$.

Thus, within the framework  of  the  above-described  method  for
correcting the canonical light-front  Hamiltonian  of  the  gauge
theory, an infinite number of counter-terms must be added to  the
Hamiltonian. Note, however, that the  formulated  conditions  for
the vanishing of the configuration are sufficient, but, generally
speaking, not necessary. Because  of  this  and  because  of  the
possible   cancellation    of    different configurations    after
integration w.r.t. transverse momenta, the  number  of  necessary
counter-terms may be smaller.

The authors are thankful to E. V. Prokhvatilov for the discussion
of the paper and for the valuable comments.

This investigation was supported by the  Russian  Foundation  for
Basic Research, Grant No. 92-02-05520-a.

\setcounter{form}{0}
\renewcommand{\theform}{A.1.\arabic{form}}

\section*{$\protect\vphantom{a}$\hfill Appendix 1}

{\bf Statement 1.} {\it If conditions (\ref{1.1}) are  satisfied,  then,  for  fixed
external momenta $p^s$ and $p^s_-\ne    0 \;\forall s$, the  equality
 \disn{A1}{
\lim_{\b\to    0}\lim_{\g\to    0}
\int\prod_k  dq_+^k  \int\limits_{V_{\scriptstyle  \e}}
\prod_k dq_-^k    {{\tilde    f(Q^i,p^s)    e^{-\g\sum_i
{Q_+^i}^2-\b
\sum_i{Q_-^i}^2} }\over{\prod_i (2Q_+^iQ_-^i-M^2_i+i\k)}}= \no
=\int\prod_k dq_+^k \int\limits_{V_{\scriptstyle \e}\cap B_L}
\prod_k dq_-^k
{{\tilde f(Q^i,p^s)}\over{\prod_i (2Q_+^iQ_-^i-M^2_i+i\k)}},
\nom}
holds while the expressions appearing in  (\ref{A1}) exist  and  the
integral  over  $\{q_+^k\}$  on  the  right-hand  side  is    absolutely
convergent. It is assumed  that  the  momenta  of  lines  $Q^i$ are
expressed in terms of loop momenta $q^k$, $V_{\e}$
is the domain corresponding to the presence of full lines,
type-1  lines,  and  type-2  lines  (the  definitions  are  given
following formula (\ref{22})), $B_L$ is the sphere of radius $L$,
where $L\ge S \: \max\limits_s |p_-^s|$, and $S$
is a number depending on the diagram structure.  }

Let us prove the statement. For each type-1 line in  (\ref{A1}),  we
perform the following partitioning:
 \dis{
\int\limits_{-\e}^{\e} dQ_-^i=\[\int dQ_-^i+(-1)
\( \int\limits_{-\infty}^{-\e}
dQ_-^i + \int\limits_{\e}^{\infty} dQ_-^i \)\].
}
Then both sides of Eq.~(\ref{A1}) become the sum of  expressions  of
the same form in which, however, the domain $V_{\e}$ corresponds to the
presence of only full and type-2  lines.  It  is  clear  that  by
proving the statement for this $V_{\e}$:  (which  is  done  below),  we
prove the original statement as well.

Let $\tilde B$ be a domain such that the surfaces on which
\hbox{$Q_-^i=0$} are not
tangent to the boundary $\tilde B$. First, we prove that in the expression
 \disn{A2}{
\int\prod_k dq_+^k
\int\limits_{V_{\scriptstyle \e}\cap \tilde B} \prod_k dq_-^k
{{\tilde f(Q^i,p^s)\: e^{-\b \sum_i{Q_-^i}^2}}\over{\prod_i
(2Q_+^iQ_-^i-M^2_i+i\k)}}
\nom}
the integral  over  $\{q_+^k\}$  is  absolutely  convergent  (here  the
integral over $\{q_-^k\}$ is finite because $\k>0$,   $\b>0$).
This  becomes
obvious (considering conditions (\ref{1.1}) and the fact that, in  type-2
lines, the momentum $Q_-^i$ is separated from zero) if  the  contours
of the integration over $\{q_-^k\}$  can be deformed in such a way  that
the momenta $Q_-^i$ of the full lines are separated from zero  by  a
finite quantity (within the domain $V_{\e}\cap \tilde  B$).
In this case, we  can
repeat the well-known Weinberg reasoning \cite{wein}.  What  can  prevent
deformation is either a "clamping" of the contour  or  the  point
$Q_-^i=0$ falling on the integration boundary.

Let us investigate the first alternative. We divide the domain of
integration over $q_+^k$  into sectors such that  the  momenta  of  all
full lines $Q_+^i$ have a constant sign  within  one  sector.  Let  us
examine one sector. We take a set of full  lines  whose  $Q_-^i$  may
simultaneously vanish. In the vicinity of  the  point  where  $Q_-^i$
from this set vanish simultaneously, we bend the contours of  the
integration over $\{q_-^k\}$ such that these contours pass  through  the
points $Q_-^i=iB^i$ and the momenta $Q_-^i$ of the type-2 lines  do  not
change. Let $B^i$ be such that $B^i  Q^i_+  \ge  0$
for the lines from  the  set
(for $Q^i_+$  from the sector under  consideration).  It  is  easy  to
check that this bending  is  possible.  (Since  the  contours  of
integration over $q_-^k$ are bent and $Q^i_-$ are expressed in terms of
$q_-^k$,
one should only check that such $b^k$ exist, where the necessary  $B^i$
are  expressed  in  the  same  way,  i.e.,  that  $B^i$  obey    the
conservation laws and flow only along the full lines). With this
bending, rather small in relation to the deviation and the  size
of the deviation region, the contours do not pass through  the
poles because, for the denominator of each line from the set  in
question, we have
 \dis{
\(2Q_+^iQ_-^i-M^2_i+i\k\)\to
\(2Q_+^i\(Q_-^i+iB^i\)-M^2_i+i\k\), \quad Q_+^iB^i\ge 0,
}
and for the other denominators, the  bending  takes  place  in  a
region separated from the point where the corresponding momenta
$Q_-^i$ are equal to $0$. Repeating the reasoning for all sets, we  can
see that there is no contour "clamping".

The other alternative is excluded by the above condition  for $\tilde B$.
To make this clear, one should introduce such coordinates $\xi^{\a}$  in
the $q^k$-space that the boundary of the domain $\tilde   B$
is  determined by the
equation $\xi^1=a=const$ and then argue  as  above  for  the
coordinates $\xi^{\a}$ with $\a\ge 2$.

After bending  the  contours,  integral  (\ref{A2}) is  absolutely
convergent in $q_+^k$, $q_-^k$ if tlie integration in $q_+^k$
is  carried  out
within the sector under consideration. On pointing out that the
result, of internal integration in (\ref{A2}) does not  depend  on
the  bending,  we  add  the  integrals  over  all  sectors    and
conclude that (\ref{A2}) converges in $\{q_+^k\}$ absolutely.

Now let us prove that if $\tilde  B$ is a quite small, finite  vicinity  of
the  point  $\{\tilde q_-^k\}$  that  lies  outside
the  sphere  $B_L$,  then
expression (\ref{A2}) is equal to zero. We  consider  the  momentum
$Q_-^i$ of one line. Flowing along
the diagram, it can ramify or it can merge  with  other  momenta.
Clearly, two  situations  are  possible:  either  it  flows  away
completely through  external  lines,  or,  probably,  after  long
wandering, part of it, $\tilde Q_-$, makes a  complete  loop.  The  former
situation is possible only if $|Q_-^i|\le \sum_r |p_-^r|$,
where all external
momenta leaving the diagram (but not  entering  it)  are  summed.
Obviously, $S$ can be chosen such that for  $\{q_-^k\}$
from  $\tilde B$,  a  line
exists whose momentum violates this condition.

The latter situation results in the existence of  a  loop,  where
the inequality $Q_-^i>\tilde    Q_-$
holds for all momenta of  its  lines  and
the positive direction of the momenta is along the loop. Then the
integral over $q_+^k$ of the loop in question can be interchanged with
the integrals over $\{q_-^k\}$ (because it is absolutely  and  uniformly
convergent for all $q_-^k$) and the residue formula  can  be  used  to
perform this integration. Since, for the loop  in  question,  the
momenta $Q_-^i$  of the lines of this loop are separated from zero and
are of the same sign, the result  is  zero.  This  has  a  simple
physical  meaning.  If  we  pass  to   stationary    noncovariant
perturbation theory, we find that only quanta  with  positive  $Q_-$
can exist. In this case, external particles with positive $p_-$  are
incoming and those with  negative  $p_-$  are  outgoing.  Then,  the
momentum conservation law favors  the  occurrence  of  the  first
situation.

The entire outside space for $\tilde B$ can  be  composed  of  the  above
domains $B_L$ (everything converges well
at infinity due to the  factor  $\exp(-\b  \sum_i{Q_-^i}^2)$).
Thus,  on  the
left-hand side of (\ref{A1}), one  can  substitute  the  integration
domain $V_{\e}\cap B_L$ for $V_{\e}$, set the limit  in  $\g$
under  the  sign  of
integration over $\{q_+^k\}$ because of its  absolute  convergence,  and
also set the limit in $\b$ under the integration sign  because  the
domain of the integration over $\{q_-^k\}$ is bounded. Thus, we  obtain
the right-hand side. The statement is proved.

{\bf Statement 2.} {\it If $V_{\e}$ corresponds  to  the  presence  of
type-2  lines alone, then, under the same conditions as in
Statement~1,  the equality
 \dis{
\int\limits_{V_{\scriptstyle \e}}\prod_k dq_-^k\int\prod_k dq_+^k
{{\tilde f(Q^i,p^s)}\over{\prod_i (2Q_+^iQ_-^i-M^2_i+i\k)}}= \cr
=\int\prod_k dq_+^k \int\limits_{V_{\scriptstyle \e}\cap B_L}
\prod_k dq_-^k
{{\tilde f(Q^i,p^s)}\over{\prod_i (2Q_+^iQ_-^i-M^2_i+i\k)}}.
}
is valid.  }

The proof of this statement is analogous to the  second
part of the proof of Statement~1.

\setcounter{form}{0}
\renewcommand{\theform}{A.2.\arabic{form}}

\section*{$\protect\vphantom{a}$\hfill Appendix 2}

{\bf Statement.} {\it If conditions (\ref{1.1})
are satisfied, the limits in  $\g$  and
$\b$ in (\ref{18}) can be interchanged (in turn)  with  the  sign  of  the
integral over $\{\a_i\}$ and then with
$\tilde f\(-i{{\dd}\over{\dd y_i}}\)$.  }

To prove this, we  define  the  vectors
\hbox{$\{q_+^1,q_-^1,\dots, q_+^l,q_-^l\}\equiv  S$},\linebreak
\hbox{$\{Q_+^1,Q_-^1,\dots,   Q_+^n,Q_-^n\}\equiv    \m    S+P$},
and \hbox{$\{y_1^+,y_1^-,\dots,  y_n^+,y_n^-\}\equiv  Y$},
where the vector $P$ is built only from external momenta and  $\m$  is
an $l\times    n$ matrix of rank $l$,
$\m^{2i}_{2k-1}=\m^{2i-1}_{2k}=0$,
$\m^{2i}_{2k}=\m^{2i-1}_{2k-1}$. Next,  we
introduce the following notation:
 \dis{
\tilde \L_i=\(\begin{array}{cc}
\g&-i\a_i \\
-i\a_i&\b
\end{array}\),\quad
\L=diag\{\tilde \L_1,\dots,\tilde \L_n\}, \quad A=\m^t\L\m, \cr
B=\m^t \L P -{1\over 2}i\m^t Y,\quad
C=-P^t\L P+iY^t P-i\sum_i\a_i M^2_i.
}
Then it follows from (\ref{19}) that
 \disn{B4}{
\hat \f(\a_i,p^s,\g,\b)=(-i)^n \tilde f\(-i{{\dd}\over{\dd y_i}}\)
\int d^{2l}S \; e^{-S^tAS-2B^tS+C}\Bigr|_{y_i=0}=\no
=(-i)^n \tilde f\(-i{{\dd}\over{\dd y_i}}\)
e^{B^tA^{-1}B+C}{{\pi^l}\over{\sqrt{\det A}}}\biggr|_{y_i=0}.
\nom}
The function $\tilde f$ is a polynomial and we consider each of its  terms
separately. Up to a factor, each term has
the form ${{\dd}\over{\dd y_{i_1}}}\dots {{\dd}\over{\dd  y_{i_r}}}$.
These derivatives act on $C$ and $B$. The action
on $C$ results in the constant factor $iN^tP$, the action on $B$
results in  the  factor
$-(1/2)iN^t\m  A^{-1}B$ or $-(1/4)N_1^t\m  A^{-1}\m^tN_2$
(the latter is the result of  the  action  of  two
derivatives; $N$, $N_1$, and $N_2$ are constant vectors).

It is necessary to prove the correctness of the  following  three
procedures: (i) setting the limit in $\g$ under the  integral  sign
for fixed $\b>0$; (ii) setting the limit in $\b$ for $\g=0$;  (iii)
setting  the  limits  in  $\g$  and  $\b$  under   the    signs    of
differentiation with respect to $Y$. In cases  (i)  and  (ii),  one
must obtain the bounds
 \disn{B5}{
|\hat \f(\a_i,p^s,\g,\b)|\le \f' (\a_i,p^s,\b),
\nom}
 \disn{B6}{
|\hat \f(\a_i,p^s,0,\b)|\le \f'' (\a_i,p^s),
\nom}
where $\f'$ and $\f''$ are functions integrable (for $\f'$ if $\b>0$)  in
any finite domain over $\a_i$, with $\a_i\ge 0$.
Then, for case (i), we have
 \dis{
|\hat \f(\a_i,p^s,\g,\b)\; e^{-\k\sum_i \a_i}|\le
\f' (\a_i,p^s,\b)\; e^{-\k\sum_i \a_i},
}
i.e., a limit on the integrated function arises, and,  thus,  the
limit in $\g$ can be put under the integral sign. The  situation  is
similar for case (ii).  It  is  evident  from  (\ref{B4})  that  the
function $\hat \f$ can be singular only if the eigenvalues of  matrix  $A$
become zero. On finding the lower bound of these eigenvalues, one
can prove through rather  long  reasoning  that  bounds  (\ref{B5}),
(\ref{B6}) exist if condition (\ref{1.1}) is satisfied.

After the limits in $\g$ and $\b$ are put under the integral sign,  it
is not difficult to interchange  them  with  the  differentiation
with respect to $Y$. One need do it only for $\a_i>0$  (for  each $i$)
and, in this case, one can  show  that  the  eigenvalues  of  the
matrix $A$ are nonzero and $\hat \f$ is not singular.

\vskip 3mm
{\noindent St.-Petersburg State University}

\end{document}